\documentclass[useAMS,usenatbib]{mn2e}

\usepackage{graphics,epsfig}
\usepackage{graphicx}
\usepackage{amssymb}

\title[GMRT observations of 3C~223.1]
{3C~223.1: A source with unusual spectral properties}
\author[D. V. Lal and A. P. Rao]{Dharam Vir Lal\thanks{E-mail:
dharam@ncra.tifr.res.in} and A. Pramesh
Rao\thanks{E-mail: pramesh@ncra.tifr.res.in} \\
\\
National Centre for Radio Astrophysics (NCRA--TIFR),
Pune University campus, Ganeshkhind, Pune - 411 007, India.}
\begin{document}

\date{Accepted 1988 December 15. Received 1988 December 14; in original form 1988 October 11}

\pagerange{\pageref{firstpage}--\pageref{lastpage}} \pubyear{2002}

\maketitle

\label{firstpage}

\begin{abstract}
Analysis of Giant Metrewave Radio Telescope low frequency data
for an $X$-shaped source, 3C~223.1 has revealed an unusual result.
The radio morphologies of it at 240 and 610 MHz show well defined
$X$-shape with a pair of active jets along the north-south axis and a
pair of wings along the east-west axis, that pass symmetrically
through the undetected radio core.
The wings (or low surface brightness jets) have flatter spectral
indices with respect to the high surface brightness jets,
which confirms the earlier marginal result obtained at high frequency
by Dennett-Thorpe et~al. (2002).
Although unusual, it is a valuable result which puts stringent
constraints on the formation models and nature of these sources.
This result clearly shows the value of mapping the sample of $X$-shaped
sources at low frequencies.
\end{abstract}


\section{Introduction}

The $X$-shaped or `winged' sources form a peculiar class of
extragalactic sources consisting of eleven sources compiled by
\citet{LeahyParma}.
These sources are characterised by two low surface brightness
lobes (the `wings') oriented at an angle to the `active', or
high surface brightness radio lobes, giving the total source
an `$X$' shape; both sets of lobes usually pass symmetrically through
the centre of the associated host galaxy.
Recently \citep{MerrittEkers} noted that seven out of eleven
are of Fanaroff-Riley type II (FR~II) \citep{FanaroffRiley}
and rest are either FR~I or mixed.

Source 3C~223.1 ($z$ = 0.1075) is a classical $X$-shaped source,
which does not reside in a rich cluster; instead it probably resides in
an environment similar to `classical' FR~IIs of similar radio power.
Although, the host galaxy seems to be a relatively undisturbed elliptical,
its {\it HST} image shows a beautiful central bulge with a dusty disk
\citep{deKoff}.
Radio observations show a high degree of polarisation
(15--30 per cent) in the
wings, and an apparent magnetic field structure parallel to the
edge of the source and along the length of the wings
\citep{Dennett}. Further high
frequency and high resolution radio polarisation
images showed field lines wrapping around the edges, as well as
complex internal structure \citep{Blacketal}.

We have started a project to study the sample of
$X$-shaped radio sources at 240 and 610~MHz using the 
Giant Metrewave Radio Telescope (GMRT).
In this paper we present results for 3C~223.1 and
describe its morphological and spectral properties.
We use the data to study  the behaviour of the
low frequency spectra at several locations (the north and south lobes
and the east and west wings) across the source
and compare it with the existing models.
We would also discuss the implications of our results
on the formation models of $X$-shaped sources.

\section{Observations}

\begin{table}
\caption{Observing log for 3C~223.1. Centre of the field
was at RA$_{\rm B1950}$ = 09:41:24.0 and Dec$_{\rm B1950}$ = 39:44:41.9.}
\centering
\begin{tabular}{l|cc}
\hline \\
 & 610 MHz & 240 MHz \\
\\
\hline \\
Observing date & 18 Dec~2003 & 18 Dec~2003 \\
Duration & 10.45 Hrs & 10.45 Hrs \\
Centre frequency & 606.68~MHz & 237.19~MHz \\
Nominal bandwidth & 16~MHz & 6~MHz \\
Effective bandwidth & 14.25~MHz & 5~MHz \\
Primary beam & 43$^{\prime}$ & 108$^{\prime}$ \\
Synthesized beam & 8$^{\prime\prime}$.6 $\times$ 5$^{\prime\prime}$.0  &
               17$^{\prime\prime}$.4 $\times$ 13$^{\prime\prime}$.3 \\
~~~~~~~~~~~~~~~(P.A.) & 80$^{\circ}$.4 & 80$^{\circ}$.3 \\
Sensitivity ($\sigma$)& 0.4~mJy~beam$^{-1}$ & 3.0~mJy~beam$^{-1}$ \\
Dynamic range & $\sim$1700 & $\sim$900 \\
Calibrator & 3C~286 & 3C~286 \\
$S_{\nu}$ (3C~286) & 21.02 Jy & 28.07 Jy \\
\\
\hline
\end{tabular}
\label{log}
\end{table}

The 240 and 610~MHz feeds of GMRT \citep{AnanthRao} are
coaxial feeds and therefore, simultaneous multi-frequency observations
at these two frequencies are possible.
We made full synthesis observations of 3C~223.1 at
240 and 610 MHz, in the dual frequency mode, using the GMRT
on 18 Dec~2003 in the standard
spectral line mode with a spectral resolution of 125 kHz.
Table~\ref{log} gives the details of the observations.
The visibility data were converted to FITS and analyzed using standard AIPS.
The flux calibrator 3C~286 was observed in the
end as an amplitude calibrator
and to estimate and correct for
the bandpass shape. We used the flux density scale which is an extension
of the \citet{Baarsetal} scale to low frequencies,
using the coefficients in AIPS task `SETJY'.
Source 0834$+$555 was used as the phase calibrator
and was observed once every 35~min.
The error in the estimated flux density,
both due to calibration and systematic, is $\lesssim$ 5\%.
The data suffered from scintillations and
intermittent radio frequency interference (RFI).
In addition to normal editing of the data, the
scintillations affected data and
channels affected due to RFI were identified and edited,
after which the central channels were averaged using AIPS task
`SPLAT' to reduce the data volume. To avoid bandwidth smearing,
5.00~MHz of clean band at 240~MHz
was reduced to 4 channels of 1.25~MHz each.
At 610~MHz where there was little RFI, 14.25~MHz of
clean band was averaged to give 3 channels of 4.75~MHz each.

While imaging, 55 facets (obtained using AIPS task `SETFC'),
spread across a $\sim$2$^\circ\times2^\circ$ field were used at 240~MHz and
12 facets covering  slightly less
than a 0.$^\circ7\times0.^\circ7$ field, were used at 610~MHz to map
each of the two fields using AIPS task `IMAGR'.
We used `uniform' weighting and the 3$-$D option for W~term
correction throughout our analysis.
The presence of a large number of point sources in the field
allowed us to do phase self-calibration to improve the image.
After 2--3 rounds of phase self-calibration, a final self-calibration
of both amplitude and phase was made to get the final image.
At each round of self-calibration, the image and the visibilities
were compared to check for the improvement in the source model.
The final maps were combined using AIPS task `FLATN' and corrected
for the primary beam of the GMRT antennas.

\begin{figure*}
\begin{center}
\begin{tabular}{l}
\includegraphics[width=16cm]{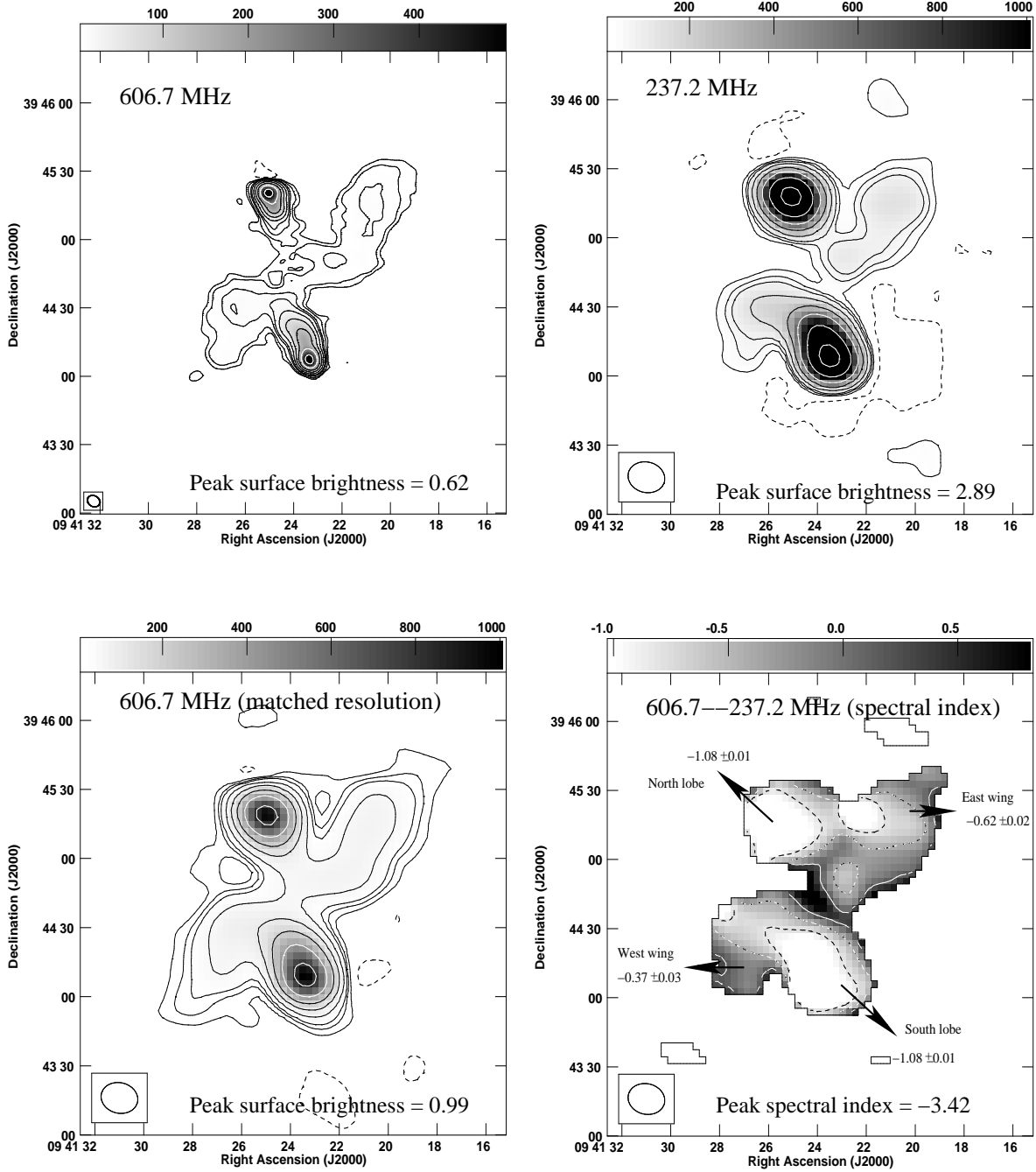}
\end{tabular}
\end{center}
\caption{Upper: Full synthesis GMRT maps of 3C 223.1 at 610
(left panel) and 240~MHz (right panel).
The CLEAN beams for 610 and 240 MHz maps are
6$^{\prime\prime}$.1~$\times$~4$^{\prime\prime}$.8
at a P.A. of 66$^{\circ}$.8
and
13$^{\prime\prime}$.1~$\times$~10$^{\prime\prime}$.8
at a P.A. of 35$^{\circ}$.2 respectively;
and the contour levels in the two maps, respectively are
$-$0.06, 0.06, 0.08, 0.16, 0.24, 0.48, 1, 2, 4, 8, 16 Jy~beam$^{-1}$
and
$-$0.08, 0.08, 0.16, 0.24, 0.48, 1, 2, 4, 8, 16 Jy~beam$^{-1}$
per cent of the peak surface brightness.
Darker regions correspond to higher surface brightness.
Lower left: The map of 3C 223.1 at 610~MHz
matched with the resolution of 240~MHz.
The contour levels are
$-$0.08, 0.08, 0.16, 0.24, 0.48, 1, 2, 4, 8, 16 Jy~beam$^{-1}$
per cent of the peak surface brightness.
Here again, the darker regions correspond to higher surface brightness.
Lower right: The distribution of the spectral index,
between 240 and 610 MHz, for the source.
The spectral index contours are spaced at 
$-$2.4, $-$1.2, 0.6 per cent of the peak spectral index.
The lighter regions represent the
relatively steep spectrum regions as compared to the darker
regions which represent flat spectrum (although the full range
of spectral index is $-$3.4 to $+$1.1, we have shown only
$-$1.0 to $+$0.8 for clarity).
The spectral indices quoted for four regions
are determined using integrated flux densities
found over a circular region of $\sim$5 pixels radius centered at the
position of tail of the arrows shown.
The error-bars on the spectral index are determined using error-bars on
the flux density found
at a source free location using similar sized circular region,
which are $\sim$3.0 and $\sim$0.4~mJy~beam$^{-1}$ at 240 and 610~MHz,
respectively.
These error-bars, both spectral indices and flux densities,
do not change significantly with increasing or decreasing the size of circle
and they also do not change significantly by changing slightly the
position of circular region.
The maximum possible systematic error in spectral index is either 0.05
across the source
or 0.14 for the wings and 0.01 for the active lobes
(see Sect.~\ref{low_f_spec}).
}
\label{full_syn}
\end{figure*}

The full synthesis radio images shown in Fig.~\ref{full_syn}
have nearly complete UV coverage, an angular resolution
$\sim$15$^{\prime\prime}$ and $\sim$8$^{\prime\prime}$ and the
rms~noise in the maps are $\sim$3.0 and
$\sim$0.4 mJy~beam$^{-1}$ at 240 and 610~MHz, respectively.
The dynamic ranges in the two maps are $\sim$900 and
$\sim$1700 respectively at 240 and 610~MHz.
The GMRT has a hybrid configuration \citep{Swarupetal}
with 14 of its 30 antennas located in a central compact array
with size $\sim$1.1~km and the remaining antennas distributed in
a roughly `Y' shaped configuration, giving a maximum baseline length
of $\sim$25~km.
The baselines obtained from antennas in the central
square are similar in length to those of the VLA~$D$-array,
while the baselines between the arm antennas are
comparable in length to the VLA~$B$-array. A single
observation with the GMRT hence yields information
on both small and large angular scales with reasonably good
sensitivity.

\section{Radio morphology}

The first high angular resolution, high sensitivity images
of 3C~223.1 at the lowest frequencies of 240~MHz
(upper right panel) and 610~MHz (upper left panel) 
are shown in Fig.~\ref{full_syn}.
This complex radio source shows an $X$-shaped morphology at both
240 and 610~MHz. The angular extent is $\sim$105~arcsec along
the apparently active lobes (those with hot-spots) and
$\sim$150~arcsec along the wings.
The nuclear source of 3C~223.1 is invisible at both these frequencies
and also in the radio maps of \citet{Dennett},
but is detected and is unresolved at 8.4~GHz \citep{Blacketal}.
The weak jet detected mid-way between core and north lobe by
\citet{Blacketal} at 8.4~GHz is not seen in our low resolution maps,
because of coarser resolution.
Our low resolution maps also suggest of a sharp boundary at the
farthest end of the north lobe and a likely ring-like feature in the
south lobe, which is consistent with earlier results of \citet{Blacketal}.

The final calibrated UV data at 610~MHz
was mapped using UV~taper of 0--22~k$\lambda$,
which is similar to that of 240~MHz data and then
restored using the restoring beam corresponding
to the 240~MHz map (Fig.~\ref{full_syn}, lower left panel).
In Table~\ref{flux_density}, we show the integrated flux densities
of 3C~223.1 along with previous measurements at other
frequencies. Our estimates at both frequencies, 240 and 610~MHz
agrees well with that of the measurements from other instruments.
We therefore believe that we have not lost any flux density in
our interferometric observations and there are no systematics
introduced in our analysis.

\begin{table}
\centering
\caption{The total intensity and spectral comparisons for source
and source regions.
The total flux densities quoted are in Jy along with corresponding error-bars
(1$\sigma$).
$\dag$: Large Cambridge interferometer \citep{Ryle};
$\parallel$: our GMRT measurements;
$\ddag$: \citet{Ficarraetal};
$\amalg$: VLA FIRST survey \citep{Beckeretal};
$\cup$: \citet{Kellermann};
$\cap$: Green Bank, Northern Sky Survey \citep{WhiteBecker}.
Last two rows show best fitted spectra to source and source regions.
$\sqcap$: Low frequency (between 240 and 610 MHz) spectral indices;
$\sqcup$: high frequency (between 1400 and 32000 MHz)
spectral indices \citep{Dennett}.
}
\begin{tabular}{rlrrrr}
\hline
\\
Freq    & \multicolumn{5}{c} {Flux density (Jy)} \\
        & \multicolumn{5}{c} {} \\ \cline {2-6}
\\
(MHz)   & Total & N lobe & W wing & S lobe & E wing \\
\\
\hline
\\
 178$^\dag$ & 8.7            & & & & \\
 240$^\parallel$ & 8.33 $\pm$0.41 & 2.95 & 0.18 & 3.27 & 0.12 \\
    &                & $\pm$0.003 & $\pm$0.003 & $\pm$0.003 & $\pm$0.003 \\
 408$^\ddag$ & 4.72 $\pm$0.10 & & & & \\
 610$^\parallel$ & 3.56 $\pm$0.17 & 1.07 & 0.10 & 1.20 & 0.09 \\
    &           & $\pm$0.001 & $\pm$0.001& $\pm$0.001 & $\pm$0.001 \\
1400$^\amalg$ & 1.90 $\pm$0.28 & 0.55 & 0.04 & 0.63 & 0.03 \\
2695$^\cup$ & 1.23 $\pm$0.61 & & & & \\
4850$^\cap$ & 0.78 $\pm$0.11 & & & & \\
\\
\hline
\\
${\alpha}^\sqcap$ &$-$0.91 $\pm$0.07 &$-$1.08 &$-$0.62 &$-$1.08 &$-$0.37 \\
                    &            &$\pm$0.01 &$\pm$0.02&$\pm$0.01&$\pm$0.03 \\
${\alpha}^\sqcup$&$-$0.75 $\pm$0.02 &$-$0.75 &$-$0.66 &$-$0.77 &$-$0.70 \\
\\
\hline
\end{tabular}
\label{flux_density}
\end{table}

\section{Low frequency radio spectra}
\label{low_f_spec}

The observations and morphology described above
allow us to investigate in detail the spectral
index distribution of 3C~223.1.
The restored and matched maps at 240 and 610~MHz,
were used further for the spectral analysis.
We determine the spectral index distribution using
the standard direct method of determining the spectral
index between maps $S_{\nu_1}(x,y)$ and $S_{\nu_2}(x,y)$
at two frequencies ${\nu_1}$ and ${\nu_2}$, given by
$$
\alpha_{\nu_1,\nu_2}(x,y) \equiv
\frac{{\rm log}~(S_{\nu_1} (x,y)/S_{\nu_2} (x,y))}{{\rm log}~(\nu_1/\nu_2)}.
$$

\begin{figure}
\begin{center}
\begin{tabular}{l}
\includegraphics[width=8.2cm]{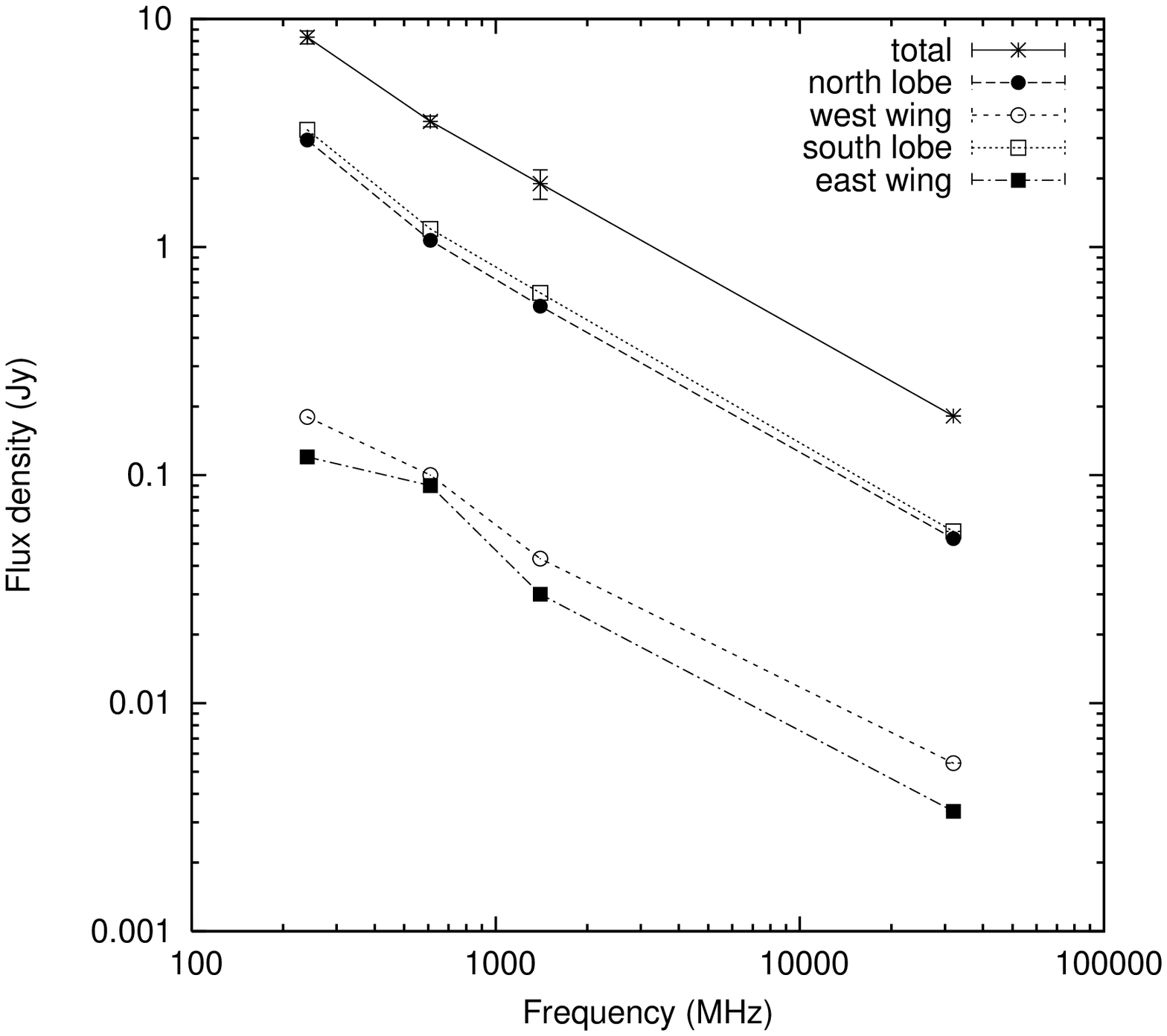}
\end{tabular}
\end{center}
\caption{Flux density for the total, the active lobes and the wings.
The 1.4~GHz measurements are from FIRST survey \citep{Beckeretal}.
The 32 GHz flux densities shown are estimated using the spectral
indices in \citet{Dennett} applied to the FIRST survey measurements.
The error-bars on the flux densities
for the source regions are $\lesssim$3.0~mJy.}
\label{flux_reg}
\end{figure}

The low frequency flux densities plotted in
Fig.~\ref{flux_reg} are calculated using the images shown in
Fig.~\ref{full_syn} (upper right and lower left panels),
which are matched to the same resolution,
and these values are tabulated in Table~\ref{flux_density}.
The flux densities at higher
frequencies in this table are quoted using
earlier observations in \citet{Dennett}.
The flux densities for the active lobes and the wings are
integrated over the region, which is at least 4 times the beam
size and  above their 3$\sigma$ contour to reduce statistical errors.
Analysis of the spectrum in different
regions of the source shows remarkable variation across the source
(Fig.~\ref{full_syn}, lower right panel).

We have obtained the best power-law fit ($S_\nu \propto \nu^\alpha$).
The low frequency (240--610~MHz) fitted
spectra have $-$0.37 $< \alpha <$ $-$1.08 for all regions
across the source.
The source also
shows evidence for steeper spectra in the active lobes than in the wings.
The east and west wings have low frequency (240--610~MHz) spectral indices,
$-$0.37 $\pm$0.03 and $-$0.62 $\pm$0.02 respectively,
whereas the north and south
active lobes have $-$1.08 $\pm$0.01 and $-$1.08 $\pm$0.01 respectively
(also see Fig.~\ref{full_syn} caption).
These results at low frequencies are consistent with those
obtained at high frequency by \citet{Dennett}, who found that
the spectral indices between 1.4 and 32~GHz to be
$-$0.70 $\pm$0.03 and $-$0.66 $\pm$0.03 for the east and west wings, and 
$-$0.75 $\pm$0.02 and $-$0.77 $\pm$0.02 for the north and south active lobes,
respectively.

This result being unusual, we were worried about the
possibility that the different UV coverages at 240 and 610 MHz
could produce some systematic errors.
While this is unlikely, since
the source is only 3--4 arcmin across and
is much smaller than the primary beam, we examined the effect
of different UV coverages by Fourier transforming the
240~MHz clean map, sampling it with the UV coverage of
610~MHz and reimaging this visibility data set.
The resultant map showed no systematic differences from the
original 240~MHz map and the rms difference in the two maps
was less than 4~\%, corresponding to the rms error in the
spectral index of $\lesssim$0.05.
Furthermore, the 240~MHz contour plot of source (Fig.~\ref{full_syn},
upper right panel) shows some evidence
that this image contains a negative depression around the source
that is less prominent at 610~MHz.
While imaging, we did not provide zero-spacing flux density, but we
have done deep cleaning, so as not to make any deconvolution errors.
Nevertheless, we look into the errors that would be introduced due to
possible negative depression. The maximum possible depression close
to the source at 240 and 610 MHz are
$-$7.6 and $-$2.7~mJy~beam$^{-1}$ respectively.
This worst case would introduce a maximum error of 0.14 in spectral
indices for the wings and 0.01 for the active lobes.
Since, the observed differences in spectral index at low frequencies
are much more than any of these uncertainties for 3C~223.1,
we believe that the observed spectral index features are real.

\section{Discussion}

Many authors have attempted to explain the unusual
structure in $X$-shaped sources. The first attempt was made by
\citet{Rees}, who suggested that the jet direction precesses
due to a realignment caused by the accretion of gas with
respect to the central black hole axis.
\citet{Dennett} discussed four possible scenarios
for the formation of such radio morphology: (1) backflow from the
active lobes into the wings \citep{LeahyWilliams,Capettietal}; (2) slow
conical precession of the jet axis \citep{Parmaetal,Macketal};
(3) reorientation of the jet axis during which flow continues; and
(4) reorientation of the jet axis, but with the jet turned off or
at greatly reduced power during the change of direction.
\citet{MerrittEkers} suggested another possible scenario,
i.e. the reorientation of black holes's spin axis due to a minor
merger, leading to a sudden flip in the direction of any associated jet.
A variant of \citet{MerrittEkers} model was suggested by
\citet{Krishna}, where the sources with $Z$ morphology within
their $X$-shapes
evolve along a $Z$-$X$ morphological sequence.
Presently, most of the observational results seem to prefer
possibilities 3, 4 of \citet{Dennett}
or \citet{MerrittEkers} or \citet{Krishna}.
The key difference between \citet{Dennett} and \citet{MerrittEkers}
reorientation models is in terms of mechanism of reorientation;
former favoured the disc instability mechanism because of
little evidence for recent merger in their sample, while the latter
preferred the coalescence scenario. Nevertheless, in these
favoured scenarios (and also in other models),
the wings are interpreted as relics of past
radio jets and the active lobes as the newer ones.

Using these radio observations,
the spectra at different locations in the source might be used
to distinguish between some of the models suggested for these sources.
In the simplest picture, the
low surface brightness wings would have an
older population of the electrons and therefore should have
steeper spectral index as compared to the
active high surface brightness radio lobes if
the magnetic fields are uniform in the measured regions.
But, the spectra of
low surface brightness jet (or the wings)
being flatter than the active high surface brightness jet,
is opposite to this simple picture.

\subsection{Interpretations based on spectral ageing}

Below we present some of the assumptions used in the
spectral ageing method for estimating the age
and discuss what could be modified to explain the above result.

1. The injection spectral index is assumed to be constant
during the active phase of the AGN and the evolution of the synchrotron
spectrum is assumed to be due to the synchrotron losses.
One of the favoured scenario for the formation of $X$-shaped sources
is the reorientation of the jet axis \citep{Dennett}.
It is possible that along with the reorientation of the jet axis
the injection spectral index also changed.
It is then possible to have a flatter spectral index in the wings
compared to the active lobes.
A similar interpretation was also suggested by \citet{Palmaetal},
to explain the anomalous spectral index features seen in NVSS~2146$+$82.

2. In classical radio galaxies, it is the usually
the high surface brightness hot-spots which consist of a young
population of energetic electrons and they are the source of plasma
for the low surface brightness lobes, which eventually
consist of older population of electrons.
In the case of $X$-shaped sources, whose formation is yet to be understood,
it is possible that this is not the case.
Instead, it is possible that the wings are in the process of becoming
new active jets, in which case it is not surprising that they have
flatter spectral index compared to the active lobes.

3. Reacceleration mechanisms are mainly seen in hot-spots
and sites undergoing interactions, which needs to be appropriately
accounted for while interpreting evolution of synchrotron spectrum.
The diffuse, low surface brightness regions
are usually believed to be relaxed systems.
It is hard to achieve reacceleration mechanisms in
the low surface brightness lobes or wings,
{\it via.} standard Alfven waves and Fermi mechanisms.
Therefore, presently it is difficult to reconcile
the observed spectral indices in the wings and the active lobes,
unless there exists an additional exotic reacceleration
mechanism.

4. Magnetic field strength does not evolve and
its distribution is believed to be isotropic,
and hence the synchrotron emission is isotropic.
\citet{Blundell} have questioned this assumption, and
have provided an alternative physical picture which could
mimic the observed spectral behaviour. They invoke
a gradient in magnetic field across the source,
together with a curved electron energy spectrum,
which would result in spectral indices being different
at distinct locations within the source and could possibly
explain our observed results.

\subsection{$X$-shaped sources--Merger of two AGNs}

We now discuss an alternative scenario, which is not
addressed earlier, for the $X$-shaped sources.

Although there are several examples of resolved
binary AGN systems, examples of unresolved binary AGN systems
are unknown. $X$-shaped sources, could be of the latter kind,
with two pairs of jets being associated with two unresolved AGNs.
An indirect hint for the presence of binary AGNs is using {\it HST} to
identify inwardly decreasing surface brightness profiles in the galaxy.
This is also found to be so using N-body simulations,
in which a coalescing binary black hole
creates a `loss cone' (displaced matter within a distance which
is roughly the separation between the two black holes when
they first form a bound pair) around it \citep{Merritt}.
We have reanalyzed the archived {\it HST} data, and
close inspection of the deconvolved brightness profile
does not suggest a centrally depressed, nearly flat core.
But this could be due to obscuration of the core by the dusty disk,
as suggested by \citet{deKoff}, and therefore we could not see
signature of the ejection of stars from the core
during the hardening of the AGNs undergoing merger.

Future high resolution, phase-referencing very long baseline
interferometry, radio imaging of
3C~223.1 could look for Keplerian orbital motion of some emission
component close to black holes, and might be used to detect the
presence of binary AGNs containing super massive black holes
\citep[e.g. 3C~66B,][]{Sudouetal}.

\section{Conclusions}

We have presented the lowest frequency image of 3C~223.1
at 240~MHz. The important consequences of these observations
combined with our 610~MHz observations are as follows:

1. The low surface brightness lobes (or the `wings') have flatter
spectra as compared to the high surface brightness active radio lobes.

2. This result is not easily explained in most models of
the formation of $X$-shaped sources.

3. Low frequency, high resolution radio images for the entire sample
of $X$-shaped sources would be useful in
understanding the formation models for these sources.
A project using GMRT is in progress and
the analysis is in advanced stage.
Our subsequent papers would present the
morphological and spectral radio results, and
discuss the results statistically
for the whole sample of $X$-shaped sources.

\section*{Acknowledgments}

We thank the staff of the GMRT who have made these observations
possible. GMRT is run by the National Centre for Radio Astrophysics
of the Tata Institute of Fundamental Research.
We also thank the anonymous referee for his/her prompt review
of the manuscript and for comments that
lead to improvement of the paper.
DVL thanks R~Nityananda, K~Subramanian and PJ~Wiita for discussions
and several useful comments and S~Ravindranath for helping with HST
data analysis.
This research has made use of the NASA/IPAC Extragalactic Database
which is operated by the Jet Propulsion Laboratory,
Caltech, under contract with the NASA, and
NASA's Astrophysics Data System.


\label{lastpage}

\end{document}